\documentstyle[prl,aps]{revtex}
\input BoxedEPS.tex
\SetOzTeXEPSFSpecial
\HideDisplacementBoxes

\begin{document}

\twocolumn[\hsize\textwidth\columnwidth\hsize
           \csname @twocolumnfalse\endcsname

\title{Identification of the bulk pairing symmetry in 
high-temperature superconductors: Evidence for an extended s-wave 
with eight line nodes}
\author{Guo-meng~Zhao} 
\address{Physik-Institut der Universit\"at Z\"urich$,$
CH-8057 Z\"urich$,$ Switzerland}

\maketitle
\widetext

\begin{abstract}
we identify the 
intrinsic bulk pairing symmetry for both electron and hole-doped cuprates 
from the existing 
bulk- and nearly bulk-sensitive experimental results such as magnetic 
penetration depth$,$ Raman scattering$,$ single-particle tunneling$,$ 
Andreev reflection$,$ nonlinear 
Meissner effect$,$ neutron scattering$,$ thermal conductivity$,$ specific 
heat$,$ and angle-resolved photoemission 
spectroscopy. These experiments 
consistently show that the dominant bulk pairing symmetry in hole-doped 
cuprates is of extended $s$-wave with eight line nodes$,$ and of 
anisotropic $s$-wave in 
electron-doped cuprates. The proposed pairing symmetries do not contradict some surface- and phase-sensitive 
experiments which show a predominant $d$-wave pairing symmetry at the degraded 
surfaces. We also quantitatively explain the phase-sensitive 
experiments along the c-axis for both Bi$_{2}$Sr$_{2}$CaCu$_{2}$O$_{8+y}$ 
and YBa$_{2}$Cu$_{3}$O$_{7-y}$.
~\\
~\\

\end{abstract}
]

\narrowtext

\section{Introduction}
An unambiguous determination of the symmetry of the order parameter 
(pair wavefunction) in cuprates is crucial to the understanding of 
the pairing mechanism of high-temperature superconductivity. 
In recent years$,$ many experiments have been designed to test the 
order parameter (OP) symmetry in the cuprate superconductors. However$,$ 
contradictory conclusions have been drawn from different experimental 
techniques 
\cite{Hardy,Jacobs,Lee,Bha,Willemin,Sacuto,Kend,Wei,Tsuei2,Tsuei3,Kirtley,Li,Sun,Ding,Kelley,Vob}$,$ which 
can be classified into being bulk-sensitive and 
surface-sensitive. For example$,$ the  magnetic penetration 
depth measurements and polarized Raman scattering experiments are bulk-sensitive. 
Angle-resolved photoemission spectroscopy (ARPES) is essentially a 
surface-sensitive technique. However$,$ the ARPES data for 
Bi$_{2}$Sr$_{2}$CaCu$_{2}$O$_{8+y}$ (BSCCO) should nearly reflect the bulk 
properties since the cleaved top surface contains an inactive Bi-O layer$,$ and 
the superconducting coherent length along the 
c-axis is very short. The single-particle tunneling experiments can probe the bulk electronic density of states when the mean free 
path is far larger than the 
thickness of the degraded surface layer \cite{Pon}. Therefore$,$ the single-particle 
tunneling experiments along the CuO$_{2}$ planes are almost 
bulk-sensitive due to a large in-plane mean free path ($>$ 100 \AA). 
In contrast$,$ the phase-sensitive experiments based on the 
Josephson tunneling are rather surface sensitive (since pair tunneling 
is limited by the coherence length which is rather short in cuprates)$,$ so that they might not 
probe the intrinsic bulk superconducting state if the surfaces are 
strongly degraded. In this case$,$ the observed product 
of the critical current times the junction normal-state resistance 
($I_{c}R_{N}$) will be very small compared with the Ambegaokar-Baratoff 
limit. Then the OP symmetry at surfaces may be different from the one 
in the bulk \cite{Bahcall}. Therefore$,$ the surface- and phase-sensitive experiments 
do not necessarily provide an acid test for the intrinsic bulk OP symmetry.

Here$,$ we identify the 
intrinsic bulk pairing symmetry for both electron and hole-doped cuprates from the existing 
bulk- and nearly bulk-sensitive experimental results such as magnetic 
penetration depth$,$ Raman scattering$,$ single-particle tunneling$,$ 
Andreev reflection$,$ nonlinear 
Meissner effect$,$ neutron scattering$,$ thermal conductivity$,$ specific 
heat$,$ and ARPES. These experiments 
consistently show that the dominant bulk pairing symmetry in hole-doped 
cuprates is of extended $s$-wave with eight line nodes$,$ and of 
anisotropic $s$-wave in 
electron-doped cuprates. The proposed pairing symmetries do not 
contradict some 
surface- and phase-sensitive 
experiments which show a $d$-wave pairing symmetry at the degraded 
surfaces. The extended $s$-wave pairing symmetry deduced from the 
bulk-sensitive experiments is also in quantitative agreeement with the 
well-designed phase-sensitive experiments along the c-axis for both Bi$_{2}$Sr$_{2}$CaCu$_{2}$O$_{8+y}$ 
and YBa$_{2}$Cu$_{3}$O$_{7-y}$.

\section{The pairing symmetry in hole-doped cuprates}

\subsection{The pairing symmetry in Bi$_{2}$Sr$_{2}$CaCu$_{2}$O$_{8+y}$}

\begin{figure}[htb]
    \ForceWidth{7cm}
	\centerline{\BoxedEPSF{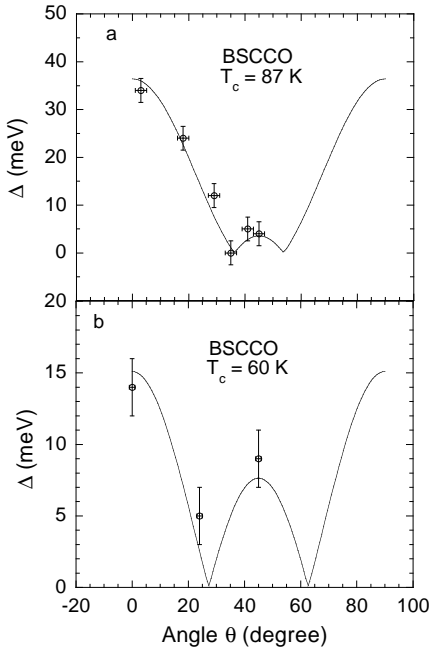}}
	\vspace{-0.3cm}
	\caption[~]{The angle 
dependence of the superconducting gap 
$\Delta (\theta )$ in the Y quadrant for 
Bi$_{2}$Sr$_{2}$CaCu$_{2}$O$_{8+y}$ (BSCCO) crystals: (a) slightly 
overdoped sample with $T_{c}$ = 87 K; (b) heavily overdoped sample 
with $T_{c}$ = 60 K. The magnitudes of the gap were extracted from 
ARPES data \cite{Ding,Vob}. Here $\theta$ is the angle measured from the 
Cu-O bonding direction. }
	\protect\label{SF1}
\end{figure}

We first examine the high-resolution ARPES data obtained for 
Bi$_{2}$Sr$_{2}$CaCu$_{2}$O$_{8+y}$ crystals \cite{Ding,Vob}. 
From the ARPES data$,$ one can determine the angle dependence of the 
superconducting gap with a resolution as high as $\pm$ 2 meV 
\cite{Vob}. Due to 
the complication arising from a possible superlattice contribution in 
the X quadrant$,$ we only use the data obtained for the Y quadrant to extract the gap function. 
In Fig.~\ref{SF1}$,$ we show the angle 
dependence of the superconducting gap 
$\Delta (\theta )$ in the Y quadrant for slightly overdoped and heavily 
overdoped BSCCO single crystals. The data were taken from 
Ref.~\cite{Ding,Vob}. Here $\theta$ is the angle measured from the 
Cu-O bonding direction. For the slightly overdoped sample (Fig.~\ref{SF1}a)$,$ the 
gap $\Delta_{D}$ at $\theta$ = 45$^{\circ}$ (diagonal direction) is very 
small (3.5$\pm$2.5 meV)$,$ and the 
gap symmetry could be consistent with a $d$-wave symmetry$,$ i.e.$,$ $\Delta (\theta ) = \Delta 
\cos 2\theta$. On the other hand$,$ the gap along the diagonal direction 
($\Gamma$-Y) for the heavily overdoped sample (Fig.~\ref{SF1}b) is not small 
(9$\pm$2 meV)$,$ which is obviously 
not consistent with the $d$-wave pairing symmetry. A similar evolution of the gap 
function with the doping has been observed by the bulk-sensitive 
polarized Raman scattering \cite{Kend}$,$ which also shows that the difference in 
the magnitudes of the gaps along the Cu-O bonding direction and the 
diagonals becomes smaller and smaller towards overdoping. 

The question is: what functional form of $\Delta (\theta )$ 
can fit the angle dependence of the gap shown in Fig.~\ref{SF1}? In general$,$ 
the gap can be expressed as $\Delta (\theta )$ = $\Delta_{s} + 
\Delta_{d}\cos 2\theta + \Delta_{g}\cos 4\theta +\ldots$. In the case 
of $\Delta_{d}$ $\simeq$ 0$,$ one has
\begin{equation}\label{Se1}
\Delta (\theta ) = \Delta (\cos 4\theta + s),
\end{equation}
where $s$ is the parameter reflecting the isotropic $s$-wave component. 
This gap function has eight line nodes for $s$ $<$ 1$,$ while 
there are no nodes for $s$ $>$ 1. The gap function (Eq.~\ref{Se1}) is 
also called extended $s$-wave (denoted by $s^{*}$-wave).  The polarized 
Raman data for an optimally doped HgBa$_{2}$CaCu$_{2}$O$_{6+y}$ are 
in good agreement with the $s^{*}$-wave gap function \cite{Sacuto}. If we 
take the absolute value of $\Delta (\theta )$$,$ then
\begin{equation}\label{Se2}
|\Delta (\theta )| = |\Delta (\cos 4\theta + s)|.
\end{equation}

We fit the data of Fig.~\ref{SF1} by Eq.~\ref{Se2}. It is 
remarkable that the fits are rather good. This indicates that the ARPES 
data may be consistent with the extended $s$-wave symmetry. The ARPES 
specified maximum gap $\Delta_{M}$ at $\theta$ = 0 for the slightly overdoped 
sample is 36$\pm$3 meV$,$ which is much larger than the value ($\sim$28 meV) 
determined from break junction spectra \cite{Miya}. On the other hand$,$ 
the ARPES determined $\Delta_{M}$ value (15$\pm$2 meV) for the heavily overdoped sample with 
$T_{c}$ = 60 K is very 
close to the value (18$\pm$2 meV) inferred from a break junction 
spectrum of a similar crystal with $T_{c}$ = 62 K \cite{DeWilde}. The discrepancy in 
the former case may be due to the fact that the doping level in the top layer 
where the ARPES probes could be slightly lower than in the bulk 
(i.e.$,$ the top CuO$_{2}$ layer might be slightly underdoped). 
Thus$,$ the ARPES experiments on the BSCCO single crystals are nearly 
bulk sensitive$,$ in contrast to the ARPES experiments on other 
cuprates$,$ which are essentially surface sensitive.

If the proposed gap functions (Eq.~\ref{Se1} and Eq.~\ref{Se2}) are indeed relevant$,$ they should 
be also consistent with other bulk-sensitive experimental results such as 
the in-plane magnetic penetration depth $\lambda_{ab} (T)$. Since there 
are eight line nodes in the proposed gap function$,$ the change of the 
in-plane penetration depth at low temperatures should be proportional to $T$. 
Following the procedure in 
Ref.\cite{Kosztin}$,$ we can readily show that the slope
\begin{equation}\label{Se3}
d\lambda_{ab} (T)/dT = [\lambda_{ab} (0)\ln 2/\Delta_{M}]\sqrt{(1 + s)/(1 - s)}.
\end{equation}
Compared with the $d$-wave symmetry$,$ the 
magnitude of the slope $d \lambda_{ab} (T)/dT$ is enhanced by a factor of  
$\sqrt{(1 + s)/(1 - s)}$.  In terms of $\Delta_{M}$ and 
$\Delta_{D}$$,$ we find that 
$s = (\Delta_{M} - \Delta_{D})/(\Delta_{M} + \Delta_{D})$ 
and $\Delta = (\Delta_{M} + \Delta_{D})/2$. Then$,$ Eq.~\ref{Se3} can be rewritten as
\begin{equation}\label{Se4}
d\lambda_{ab} (T)/dT = \lambda_{ab} (0)\ln 2/\sqrt{\Delta_{M}\Delta_{D}}.
\end{equation}
It is interesting to see that $d \lambda_{ab} (T)/[\lambda_{ab} (0) dT]$ is 
inversely 
proportional to $\sqrt{\Delta_{M}\Delta_{D}}$$,$ namely$,$ the geometric 
average of $\Delta_{M}$ and $\Delta_{D}$.

\begin{figure}[htb]
    \ForceWidth{7cm}
	\centerline{\BoxedEPSF{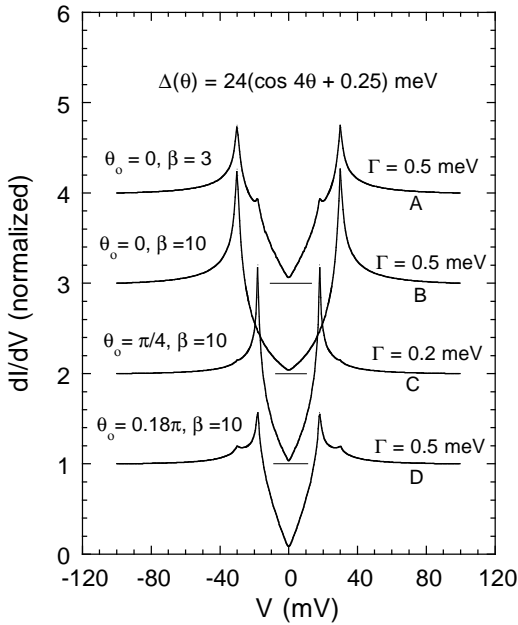}}
	\vspace{0.3cm}
	\caption[~]{Numerically calculated curves of the renormalized $dI/dV$ for 
	a gap function of $\Delta (\theta ) = \Delta (\cos 4\theta + 
s)$ with $\Delta$ = 24 meV and $s$ = 0.25. The four curves correspond 
to different values of the parameters $\Gamma$$,$ $\beta$$,$ and 
$\theta_{o}$$,$ which are indicated in the figure. The curves A$,$ B and 
C are vertically shifted up by 3$,$ 2 and 1$,$ respectively.}
	\protect\label{SF2}
\end{figure}

The single-particle tunneling spectroscopy can probe the superconducting density of 
states (DOS) with fine energy resolution and considerable 
directionality. For an isotropic $s$-wave superconductor$,$ the 
characteristic $dI/dV$ 
vs $V$ curve in the point-contact SIN tunneling junctions 
exhibits a step-like peak at a voltage 
$V_{p}$ = $\Delta/e$. For an anisotropic gap function $\Delta (\theta)$$,$ 
the directional dependence of the tunneling differential conduction is 
given by \cite{Suzuki}
\begin{equation}\label{Se5}
\frac{dI}{dV} \propto \int_{0}^{2\pi}p(\theta 
-\theta_{0}) \Re [\frac{eV - i\Gamma}{\sqrt{(eV - i\Gamma)^{2} - 
\Delta^{2}(\theta)}}]N(\theta) d\theta.
\end{equation}
Here $N(\theta)$ represents the anisotropy of the band dispersion; 
$\Gamma$ is the life-time broadening parameter of an electron;
$p(\theta -\theta_{0})$ is the angle dependence of the 
tunneling probability which decays exponentially as 
$p(\theta -\theta_{0})$ = $\exp [-\beta \sin^{2}(\theta 
-\theta_{0})]$ ($\theta_{0}$ is the angle of the tunneling 
barrier direction); the parameter $\beta$ decreases with decreasing barrier resistance $R_{N}$. 
For simplicity$,$ we assume a cylindrical 
Fermi surface$,$ so that both $N(\theta)$ and $\beta$ are independent of 
the angle. This will not change the basic features of the 
$dI/dV$ curve. In Fig.~\ref{SF2} we show the 
numerically calculated results of the renormalized $dI/dV$ for a gap 
function of $\Delta (\theta ) = \Delta (\cos 4\theta + 
s)$ with $\Delta$ = 
24 meV and $s$ = 0.25. One can readily show that the maximum gap is $\Delta_{M} = (1 + 
s) \Delta$ = 30 meV at 
$\theta$ = 0$,$ and the gap along the diagonal directions is 
$\Delta_{D} = (1 - s)\Delta$ = 18 meV. From Fig.~\ref{SF2}$,$ one can 
see that either two or four peak features appear clearly in the 
$dI/dV$ curves$,$ depending on the tunneling 
barrier direction and/or the $\beta$ value. For a small $\beta$ value
(corresponding to a small barrier resistance)$,$ four peak features are well defined (see curve A). 
The peak positions are located at $eV$ = 
$\pm\Delta_{M}$ and $\pm\Delta_{D}$. Therefore$,$ from the 
peak positions$,$ we can determine $\Delta_{M}$ and 
$\Delta_{D}$.

\begin{figure}[htb]
   \ForceWidth{7cm}
	\centerline{\BoxedEPSF{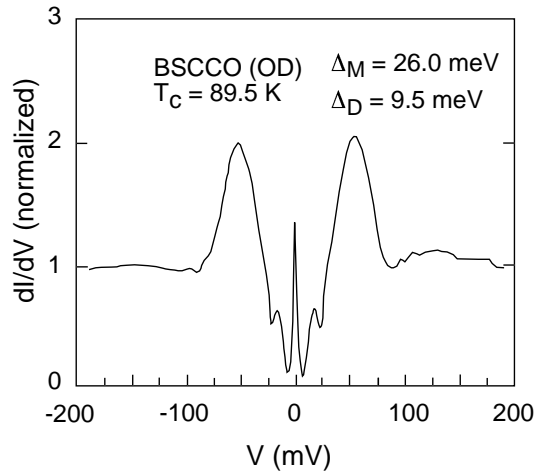}}
	\vspace{0.3cm}
	\caption[~]{Normalized $dI/dV$ curves at 14 K for the SIS break 
junctions on a slightly overdoped BSCCO crystal. The spectra  were taken  
from Ref.~\cite{Mour}.}
	\protect\label{SF3}
\end{figure}

In Fig.~\ref{SF3}$,$ we plot the normalized $dI/dV$ curve at 14 K for an SIS break 
junction on a BSCCO crystal which is slightly overdoped ($T_{c}$ = 90 K) \cite{Mour}. The 
junction has a very low barrier resistance ($\sim$200 $\Omega$)\cite{Mour}$,$ 
indicating a small $\beta$ value. It is remarkable that there are four well-defined 
peak features in the spectrum$,$ which resemble curve A in Fig.~\ref{SF2}. 
The pronounced zero bais peak arises from Josephson tunneling 
\cite{Miya,Mour}. When 
the barrier resistance is above 2 k$\Omega$$,$ the inner gap features 
disappear \cite{Mour}$,$ in agreement with curve B in Fig.~\ref{SF2}. We would like to 
mention that$,$ for SIS break junctions$,$ the peak positions are located at $eV$ = 
$\pm 2\Delta_{M}$ and $\pm 2\Delta_{D}$. From the spectra$,$ we obtain 
$\Delta_{M}$ = 26$\pm$0.5 meV$,$ and $\Delta_{D}$ = 9.5$\pm$0.5 meV. 
The $\Delta_{M}$ value obtained from the break junction spectrum is the 
same as that found from the c-axis intrinsic tunneling junctions made 
of the insulating Bi-O layers \cite{PMuller}. From the $\Delta_{M}$ and $\Delta_{D}$ values$,$ we deduce a gap 
function $\Delta (\theta ) = \Delta (\cos 4\theta 
+ s)$ with $\Delta$ = 
17.75 meV and $s$ = 0.46. With this gap function and $\lambda_{ab}
(0) $ = 2690$\pm$150 \AA~\cite{Proz}$,$ we calculate from Eq.~4 that 
$d\lambda_{ab} (T)/dT$ = 10.2$\pm$0.6 \AA/K$,$ in excellent agreement 
with the measured values (10.2$\pm$0.2 \AA/K) \cite{Jacobs,Lee}. Similarly$,$ 
the earlier break junction 
spectra for an overdoped BSCCO with $T_{c}$ = 86 K also indicate 
double gap features at $\Delta_{M}$ = 24$\pm$2 meV and at $\Delta_{D}$ = 
12$\pm$1 meV (Ref.~\cite{Bus}). The tunneling spectra are in good agreement with ARPES data for an 
overdoped BSCCO with $T_{c}$ = 83 K \cite{Ma}. The ARPES experiment clearly 
showed that $\Delta_{M}$ = 20$\pm$2 meV and $\Delta_{D}$ = 12$\pm$2 
meV \cite{Ma}.  
Moreover$,$ the 
inner gap features also appear in SIS break junction 
spectra of a heavily overdoped crystal with $T_{c}$ = 62 K$,$ corresponding 
to $\Delta_{D}$ = 7.5-9.0 meV (Ref.~\cite{DeWilde,Ozyuzer}). The magnitude of $\Delta_{D}$ is in 
excellent agreement 
with that found from the ARPES experiment (see Fig.~\ref{SF1}b).

We would like to point out that the values of $\Delta_{M}$ determined 
from Raman spectrum of $B_{1g}$ symmetry may be overestimated due to 
the fact that the extended van Hove singularity is slightly below 
the Fermi level. In this case$,$ the spectra would show double 
peaks at Raman shifts of 2$\Delta_{M}$ and 2$\sqrt{\Delta^{2}_{M} +\xi_{vH}^{2}}$$,$ 
where $\xi_{vH}$ is the energy position of the van Hove singularity below the Fermi level. When 
$\xi_{vH}$$<$$<$ $\Delta_{M}$$,$ one can only see a single broad peak 
slightly below 2$\sqrt{\Delta^{2}_{M} + \xi_{vH}^{2}}$.

\subsection{The pairing symmetry in YBa$_{2}$Cu$_{3}$O$_{7-y}$}

\begin{figure}[htb]
\vspace{0.2cm}
    \ForceWidth{7cm}
	\centerline{\BoxedEPSF{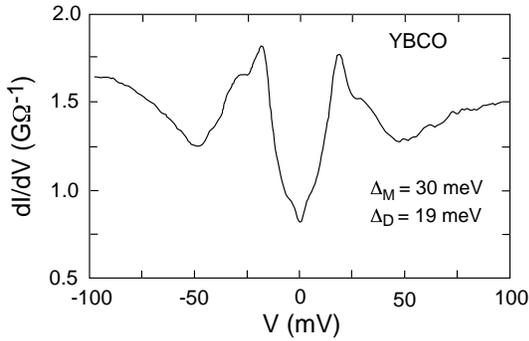}}
	\vspace{0.2cm}
	\caption[~]{Scanning tunneling spectrum for a slightly overdoped 
YBa$_{2}$Cu$_{3}$O$_{7}$ (YBCO) crystal. The spectrum was taken from 
Ref.~ \cite{Mag}.}
	\protect\label{SF4}
\end{figure}

Evidence for an extended $s$-wave pairing symmetry 
in YBa$_{2}$Cu$_{3}$O$_{7-y}$ (YBCO) also comes from single-particle tunneling spectra. Fig.~\ref{SF4} shows 
scanning tunneling spectrum for a slightly overdoped 
YBa$_{2}$Cu$_{3}$O$_{7-y}$ crystal \cite{Mag}.  Four peak features appear in 
this spectrum that is 
similar to curve D in Fig.~\ref{SF2}. From the peak positions$,$ we obtain 
$\Delta_{M}$ = 30$\pm$2 meV$,$ and $\Delta_{D}$ = 19$\pm$1 meV. The size of 
$\Delta_{M}$ $\simeq$ 30 meV is consistent with a break junction spectrum 
\cite{Pon}$,$ 
and a scanning tunneling spectrum along the a-axis direction 
\cite{Wei}. A gap 
feature with $\Delta_{D}$ = 19 meV was also seen in a scanning 
tunneling spectrum \cite{Wei} that is very similar to curve C in 
Fig.~\ref{SF2}.

Now we discuss the Andreev reflection. Since there is sign change 
about its nodal directions in our extended $s$-wave order parameter$,$ the 
Andreev-bound surface states can be formed. This will lead to a zero-bias 
conduction peak (ZBCP) if tunneling is nearly along one of 
the nodal directions and the bare Fermi velocities between the cuprates and 
normal metals (e.g.$,$ Ag and Au) are well matched. For hole doped 
cuprates$,$ the bare Fermi velocity $v^{b}_{F}$ strongly depend on the angle $\theta$$,$ that is$,$ $v^{b}_{F}$ is 
small along the 
bonding direction$,$ and large along the diagonal directions. This 
implies that the observation of the Andreev reflection is difficult for 
tunneling along the bonding direction since the value 
of $v^{b}_{F}$ along this direction is small compared with that of Au or Ag. Due to the strong 
anisotropy of $v^{b}_{F}$ in cuprates$,$ the Andreev reflection mainly 
probes the gap feature at $eV$ = $\Delta_{D}$. If tunneling is 
along one of the diagonal directions$,$ and the angle between 
the nodel and diagonal directions is far larger than the half tunneling 
angle (depending on $\beta$)$,$ one can see an $s$-wave like gap 
approximately equal 
to $\Delta_{D}$ in the Andreev reflection spectra. Indeed an $s$-wave 
like gap feature at $eV$ $\simeq$ 20 meV has been observed in the Andreev 
reflection spectra of several YBCO crystals with $T_{c}$ = 90 K 
\cite{Yagil}. We 
would like to mention that$,$ in general$,$ the double gap features should 
also appear in the Andreev reflection spectra when the $\beta$ value is 
small and $v^{b}_{F}$ does not have a significant anisotropy.

The tunneling data of YBCO (Fig.~\ref{SF4}) are thus consistent with a gap 
function $\Delta (\theta ) = \Delta (\cos 4\theta + s)$ with $\Delta$ = 
24.5 meV and $s$ = 0.225. This gap function is in quantitative agreement with 
the a-axis $\lambda_{a} (T)$ data (which reflect magnetic screening in CuO$_{2}$ planes) for a fully oxygenated YBCO crystal 
\cite{Kamal}. From Eq.~\ref{Se4}$,$ we calculate  
$d\lambda_{a} (T)/dT$ = 4.0 \AA/K~using 
$\lambda_{a} (0)$ = 1600 \AA~ (Ref.~\cite{Kamal})$,$ $\Delta_{D}$ = 19 meV$,$ and 
$\Delta_{M}$ = 30 meV. We will get the same value of $d\lambda_{a} (T)/dT$ 
if we use $\Delta_{D}$ = 21 meV$,$ and 
$\Delta_{M}$ = 27 meV. For a 
$d$-wave gap function $\Delta(\theta) = \Delta_{M}\cos 2\theta$ with 
$\Delta_{M}$ = 30 meV$,$ the calculated $d\lambda_{a} (T)/dT$ = 3.2 
\AA/K. The measured 
value of $d\lambda_{a} (T)/dT$ is 4 \AA/K \cite{Kamal}. It is evident 
that the extended $s$-wave gap function is in much better agreement 
with experiment than the $d$-wave gap function.
\begin{figure}[htb]
    \ForceWidth{7cm}
	\centerline{\BoxedEPSF{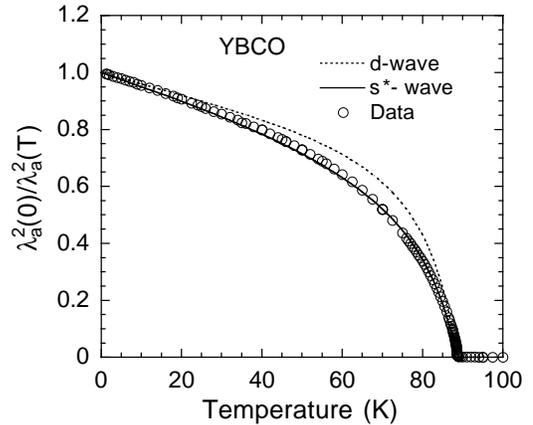}}
	\vspace{0.3cm}
	\caption[~]{Temperature dependence of 
the a-axis $\lambda_{a}^{2}(0)/\lambda_{a}^{2}(T)$ for a very high-quality 
YBa$_{2}$Cu$_{3}$O$_{7}$ (YBCO) crystal with $T_{c}$ = 88.7 K. The 
data 
were taken from 
Ref.~ \cite{Kamal}. The solid line is the calculated curve for the 
$s^{*}$-wave gap function deduced from the tunneling spectrum in 
Fig.~\ref{SF4}. The dash line is the calculated curve for a $d$-wave gap 
function with $\Delta_{M}$ = 30 meV. }
	\protect\label{SF5}
\end{figure}
Now we calculate the temperature dependence of 
$\lambda_{ab}^{2}(0)/\lambda_{ab}^{2}(T)$ for the $s^{*}$-wave gap function. For a cylindrical 
Fermi surface \cite{Jacobs}
\begin{equation}\label{Se6}
\frac{\lambda_{ab}^{2}(0)}{\lambda_{ab}^{2}(T)} = 1 + 
(1/\pi)\int_{0}^{2\pi}\int_{0}^{\infty}d\theta d\epsilon 
\frac{\partial f}{\partial E}.
\end{equation}
Here $E = \sqrt{\epsilon^{2}+ \Delta^{2}(\theta,T)}$; $f$ is the 
Fermi-Dirac distribution function; $\Delta(\theta,T) = \Delta 
(T)(\cos 4\theta + 
s)$; $\Delta (T)$ = $\Delta \tanh (2.2\sqrt{T/T_{c} -1})$ 
(Ref.~\cite{Thelen}). In  Fig.~\ref{SF5}$,$ 
we compare the experimental data for YBCO (open circles) \cite{Kamal} and the numerically 
calculated result (solid line) for the above deduced gap function $\Delta (\theta ) = 
24.5 (\cos 4\theta + 0.225)$ meV. It is remarkable that the data are 
in quantitative agreement with the calculated result without any 
fitting parameters. The dash line is the calculated result for a 
$d$-wave gap function $\Delta(\theta) = \Delta_{M}\cos 2\theta$ with 
$\Delta_{M}$ = 30 meV. It is clear that the agreement between the data and 
the calculated curve is poor for the $d$-wave symmetry. It is worthy to 
note that the temperature dependence of 
$\lambda_{ab}^{2}(0)/\lambda_{ab}^{2}(T)$ is mainly determined by the 
gap function$,$ so the shape of the Fermi surface has little effect 
on $\lambda_{ab}^{2}(0)/\lambda_{ab}^{2}(T)$.

\begin{figure}[htb]
    \ForceWidth{7cm}
	\centerline{\BoxedEPSF{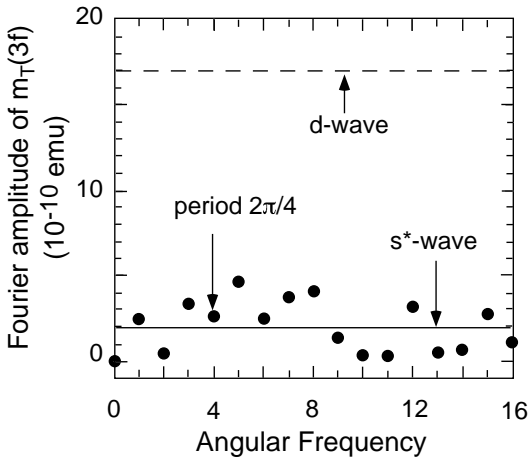}}
	\vspace{0.3cm}
	\caption[~]{Sine Fourier amplitudes of the transverse 
magnetization $m_{T}$ in the Meissner state for a high-quality 
YBa$_{2}$Cu$_{3}$O$_{7}$ (YBCO) crystal. The data 
were taken from 
Ref.~ \cite{Bha}. The solid line is the predicted sine Fourier 
amplitude at $2\pi/4$ for the 
$s^{*}$-wave gap function deduced from the tunneling spectrum in 
Fig.~\ref{SF4} and the a-axis $\lambda_{a} (T)$  data in Fig.~\ref{SF5}. The calculated amplitudes at 
$2\pi/2$$,$ $2\pi/3$$,$ $2\pi/4$$,$ and 
$2\pi/5$ are similar while the ones at other periods are much smaller. The dash line is the 
predicted sine Fourier amplitude at $2\pi/4$ for a $d$-wave gap function.}
	\protect\label{SF6}
\end{figure}

The gap function of YBCO deduced from the tunneling and the 
$\lambda_{a} (T)$ data is also consistent with the measured transverse 
magnetization $m_{T}$ in the Meissner state \cite{Bha}$,$ as plotted in 
Fig.~\ref{SF6}. This bulk sensitive 
experiment shows a very small sine fourfold component of the transverse 
magnetization$,$ which is at 
least 4 times smaller than the predicted value from the $d$-wave 
symmetry. This indicates that the dominant pairing symmetry is not 
the $d$-wave. Using the formulas reported in Ref. \cite{Zutic}$,$ we can calculate 
the sine components of 
the transverse 
magnetization for the $s^{*}$-wave gap function deduced above. We find that the 
sine fourfold component for the $s^{*}$-wave OP is a 
factor of 8.9 smaller than for the pure $d$-wave OP. The predicted 
sine Fourier amplitude at period  $2\pi/4$ is indicated by a horizontal solid 
line in Fig.~\ref{SF6}. The calculated amplitudes at $2\pi/2$$,$ $2\pi/3$$,$ $2\pi/4$$,$ and 
$2\pi/5$ are similar while the ones at other periods are much smaller. It is clear that the 
predicted amplitudes at all the periods are below the noise level which is 
about 5$\times$10$^{-10}$ emu \cite{Bha}. Therefore$,$ the very
small nonlinear Meissner effect observed in the overdoped YBCO is in agreement with 
the $s^{*}$-wave OP or with a nodeless OP \cite{Bha} rather than with the 
$d$-wave OP. A nodeless OP symmetry is in contradiction with the 
observed  linear $T$ 
dependence of the thermal conductivity down to a very low 
temperature (50 mK) \cite{Chiao}.

In addition$,$ we further show that the $s^{*}$-wave gap function 
is in quantitative agreement with the low temperature thermal conductivity$,$ 
specific heat$,$ and surface Andreev bound states. 
By replacing $\Delta_{M}$ by $\sqrt{\Delta_{M}\Delta_{D}}$ in the 
equations for the low-temperature electronic thermal 
conductivity $\kappa_{el}$ and specific heat $C_{el}$ for the $d$-wave 
gap function in the clean limit \cite{Chiao}$,$ we obtain the following equations for an 
$s^{*}$-wave gap function:
\begin{equation}\label{Se4a}
\frac{\kappa_{el}}{T} = 
\frac{k_{B}^{2}v_{F}k_{F}}{6d\sqrt{\Delta_{M}\Delta_{D}}},
\end{equation}
and
\begin{equation}\label{Se5a}
\frac{C_{el}}{T^{2}} = \frac{9\zeta (3)k_{B}^{3}k_{F}}{\pi\hbar 
v_{F}d\sqrt{\Delta_{M}\Delta_{D}}}.
\end{equation}
Here $v_{F}$ and $k_{F}$ are the Fermi velocity and momentum along the nodal 
directions$,$ respectively; $d$ is the average interlayer distance; $\zeta$(3) = 1.20. 
One should note that impurity scattering tends to suppress the values of both $\frac{\kappa_{el}}{T}$ and  
$\frac{C_{el}}{T^{2}}$. The Fermi velocity along the nodal 
directions has recently been obtained for YBCO from the studies of surface 
Andreev bound states \cite{Carrington2001}. The deduced Fermi velocity 
$v_{F}$ is (1.2$\pm$0.2)$\times$10$^{5}$ m/s$,$ which is a factor of 2 
smaller than the measured Fermi velocity along the diagonal 
directions from the ARPES data of BSCCO \cite{Kaminski}. This suggests that the nodal 
directions might be far away from the diagonal directions. For the 
$s^{*}$-wave gap function deduced above for overdoped YBCO$,$ the nodal 
directions are about 19$^{\circ}$ away from the diagonal directions 
(i.e.$,$ at $\theta$ = 26$^{\circ}$). Indeed$,$ from 
the ARPES data of BSCCO \cite{Kaminski}$,$ one can clearly see that the Fermi velocity 
at $\theta$ = 26$^{\circ}$ is smaller than that at $\theta$ = 
45$^{\circ}$ by a factor of about 2. Substituting $v_{F}$ = 1.2$\times$10$^{5}$ m/s$,$ 
$k_{F}$ = 0.7 \AA$^{-1}$ \cite{Kaminski}$,$ $d$ = 5.85 \AA$,$ 
$\Delta_{M}$ = 30 meV and $\Delta_{D}$ = 19 meV into Eq.~\ref{Se4a} and 
Eq.~\ref{Se5a}$,$ we 
obtain $\frac{\kappa_{el}}{T}$ = 0.12 mW/K$^{2}$cm and  
$\frac{C_{el}}{T^{2}}$ = 0.24 mJ/moleK$^{3}$. The calculated values are in 
excellent agreement with the measured values: $\frac{\kappa_{el}}{T}$ = 
0.14$\pm$0.03 mW/K$^{2}$cm (Ref.~\cite{Chiao}) and  
$\frac{C_{el}}{T^{2}}$ = 0.20$\pm$0.05 mJ/moleK$^{3}$ (Ref.~\cite{Junod}).

Moreover$,$ thermal conductivity of YBCO as a function of angle of an 
inplane magnetic field relative to the crystal axes has been studied 
both theoretically and experimentally \cite{Yu95,Aubin}. A theoretical calculation 
for the angular dependence of the magnetothermal conductivity 
\cite{Yu95} 
shows that an extended $s$-wave gap produces a more symmetric angular variation 
than a $d$-wave gap. It appears that both sets of experimental data 
\cite{Yu95,Aubin} are more consistent with an extended $s$-wave gap than 
a $d$-wave gap.

\subsection{The pairing symmetry in La$_{2-x}$Sr$_{x}$CuO$_{4}$}

The polarized 
Raman scattering data \cite{Chen} for nearly optimally-doped 
La$_{2-x}$Sr$_{x}$CuO$_{4}$ (LSCO) with $T_{c}$ = 37 K yield 
2$\Delta_{M}/k_{B}T_{c}$ = 7.7. From the measured value of $d\lambda_{ab}
(T)/[\lambda_{ab} (0) dT$] for the optimally-doped LSCO 
\cite{Panagopoulos}$,$ one can readily 
calculate  2$\sqrt{\Delta_{M}\Delta_{D}}/k_{B}T_{c}$ = 4.2 using 
Eq.~\ref{Se4}. Then we get 2$\Delta_{D}/k_{B}T_{c}$ = 2.3$,$ i.e.$,$ 
$\Delta_{D}$ = 3.8 meV. This value is in good agreement with the Andreev 
reflection spectrum of optimally-doped LSCO \cite{Deut}$,$ which shows the $s$-wave 
like gap feature at $eV$ $\simeq$ 3.5 meV. Therefore$,$ three independent 
bulk-sensitive experiments on the optimally-doped LSCO consistently suggest 
a gap function: $\Delta (\theta ) = 
8.1(\cos 4\theta + 0.53)$ meV with $\Delta_{D}$ = 3.8 meV and 
$\Delta_{M}$ = 12.5 meV.

Now we can quantitatively explain the neutron scattering experiment on 
an optimally-doped LSCO single crystal \cite{Lake}. The experiment shows that low 
energy magnetic excitations are peaked at the quartet of wavevectors 
(0.5$\pm$0.135$,$ 0.5) and (0.5$,$ 0.5$\pm$0.135) in the normal state$,$ 
and a spin gap with energy of about 6.7 meV appears in the low-temperature 
superconducting state. The magnitude of the spin gap should be equal 
to twice the superconducting gap along the incommensurate wave vectors 
(i.e.$,$ at $\theta$ = 39$^{\circ}$) \cite{Mason}. From the gap function deduced 
above$,$ we calculate $2\Delta (39^{\circ})$ = 6.2 meV$,$ in remarkably 
good agreement with experiment. Moreover$,$ it was also found 
\cite{Lake} that the 
spin gap at  $\theta$ = 45$^{\circ}$ is 6$\pm$2 meV$,$ which is 
consistent with $2\Delta_{D}$ = 7.6 meV within the experimental 
uncertainty. Obviously$,$ the $d$-wave gap function is incompatible with the 
large spin gap observed along the diagonal direction. The neutron 
data might be also consistent with an isotropic spin gap$,$ as suggested 
by Lake {\em et al.} \cite{Lake}. However$,$ the isotropic spin gap is 
incompatible with the $T^{3}$ dependence of the spin-lattice 
relaxation rate observed in hole-doped cuprates. Only with the 
$s^{*}$-wave gap function for LSCO$,$ one can quantitatively explain the neutron 
experiment$,$ Raman scattering$,$ magnetic penetration depth$,$ Andreev 
reflection$,$ and magnetic resonances.

\section{The pairing symmetry in electron-doped cuprates}

The recent measurements 
of $\lambda_{ab} 
(T)$ in an electron-doped  Pr$_{1.85}$Ce$_{0.15}$CuO$_{4-y}$ (PCCO) 
reveal contradictory results \cite{Alff,Prozorov}.  In a high-quality 
PCCO thin film with the lowest residual resistivity and the 
highest $T_{c}$$,$ the temperature 
dependence of $[\lambda_{ab} (T) -\lambda_{ab} (0)]/\lambda_{ab} (0)$ is consistent with an $s$-wave pairing 
symmetry with a reduced energy gap $2\Delta (0)/k_{B}T_{c}$ = 2.9 
\cite{Alff}. On 
the other hand$,$  the low-temperature $\lambda_{ab} (T)$ in less ideal 
PCCO single crystals exhibits a power-law temperature dependence$,$ as expected from a 
dirty $d$-wave superconductor \cite{Prozorov}. 

We show that these apparently conflicting data might 
well be reconciled by a deeper 
understanding of how microstructure affects screening. It is well 
known$,$ for example$,$ that the screening length in weakly coupled 
Josephson array of grains is dominated by the magnitude and 
temperature dependence of the Josephson coupling current between array 
elements \cite{Gio}. Thus$,$ tunnel coupling across grain boundaries 
and/or planar defects (weak links)$,$ rather than 
the BCS response of the grains themselves$,$ mainly determines the 
magnetic screening length$,$ surface resistance$,$ and critical current 
(see a review article \cite{Halbreview}).   The extrinsic 
effect due to the weak links can lead to a linear $T$ dependence in 
the effective $\lambda_{ab} (T)$ at low temperatures and to a large residual 
surface resistance \cite{HalbJAP}. Similarly$,$ Hebard {\em et al.} \cite{Hebard} 
showed that the current-induced 
nucleation of vortex-antivortex pairs at defects can make an additional extrinsic 
contribution to the screening length$,$ i.e.$,$ a pinning penetration 
depth $\lambda_{ab}^{p}(T)$. Within this scenario$,$ the  $\lambda_{ab}^{p}(T)$ 
in zero magnetic field is 
given by \cite{Hebard}
\begin{equation}\label{Se7}
\lambda_{ab}^{p}(t) = \lambda_{ab}^{p}(0)/(1 - t^{2}),
\end{equation} 
where $t = T/T_{c}$; $\lambda_{ab}^{p}(0)$ = 
$[\Phi_{0}/H_{c}(0)]\sqrt{2N_{d}/\pi}$; $\Phi_{0}$ is the 
flux quantum; $N_{d}$ is the areal density of uniformly distributed 
defects; $H_{c}(0)$ is the zero-temperature critical field. In the 
presence of the external dc field $H$$,$ the expression for 
$\lambda_{ab}^{p}(0,H)$ has to be modified \cite{HalbJAP}. The total 
screening length is $\lambda_{ab} (t) = \sqrt{[\lambda^{L}_{ab}(t)]^{2} + 
[\lambda_{ab}^{p}(t)]^{2}}$$,$ where $\lambda^{L}_{ab}(t)$ is the intrinsic London 
penetration depth \cite{Hebard}. Assuming an $s$-wave pairing symmetry$,$ 
we readily show that the $\lambda_{ab} (T)$ 
at low temperatures (below 0.2$\Delta (0)/k_{B}$) is given by
\begin{eqnarray}\label{Se8}
\lambda_{ab} (T) = \lambda_{ab} (0) + 
\frac{[\lambda^{L}_{ab}(0)]^{2}} 
{\lambda_{ab} (0)}\sqrt{\pi\Delta (0)
/2k_{B}T} \nonumber\\
\exp [-\Delta (0)/k_{B}T]
+\frac{\lambda_{ab}^{2}(0) - 
[\lambda^{L}_{ab}(0)]^{2}}{\lambda_{ab} (0)T^{2}_{c}}T^{2}.
\end{eqnarray}

It is clear that the $T^{2}$ dependence of $\lambda_{ab} (T)$ 
at low temperatures in zero field can be completely caused by the extrinsic effect$,$ 
that is$,$ the nucleation of vortex-antivortex pairs at defects. If $N_{d}$ is 
negligible$,$ $\lambda_{ab} (0) = \lambda^{L}_{ab}(0)$$,$ and the second term in 
Eq.~\ref{Se8} is absent. Then we recover the BCS expression \cite{Muh}$,$
\begin{equation}\label{Se9} 
\lambda_{ab} (T) = \lambda_{ab} (0) + \lambda_{ab} (0)\sqrt{\pi\Delta (0)/2k_{B}T}\exp[-\Delta 
(0)/k_{B}T].
\end{equation}

\begin{figure}[htb]
    \ForceWidth{7cm}
	\centerline{\BoxedEPSF{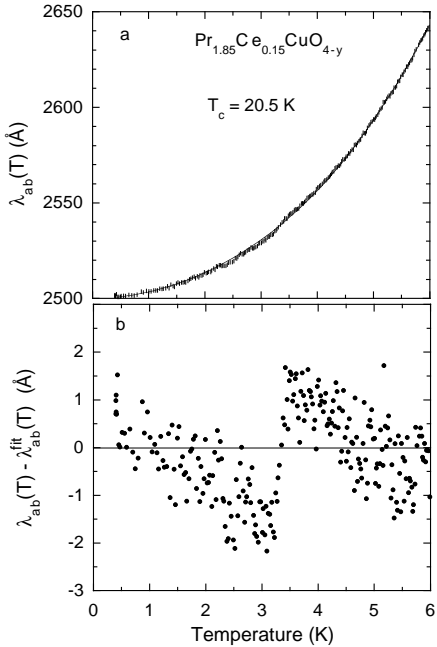}}
	\vspace{-0.6cm}
	\caption[~]{(a) Temperature dependence of $\lambda_{ab} (T)$ below 6 
K for a PCCO single crystal. The solid line is the fitted curve by 
Eq.~\ref{Se8} with $2\Delta (0)/k_{B}T_{c}$ = 2.9 and 
$\lambda_{ab}^{L}(0)$ = 1643 \AA. The value of $\lambda_{ab}^{L}(0)$ was found 
to be 1600$\pm$100 \AA~from the optical data \cite{Homes}. (b) The 
difference between the data and the 
fitted curve. The data are from Ref.~\cite{Prozorov}. }
	\protect\label{SF7}
\end{figure}

In Fig.~\ref{SF7}a$,$ we plot temperature dependence of $\lambda_{ab} (T)$ below 6 
K for a PCCO single crystal (the data are from Ref.~\cite{Prozorov}). 
The zero-temperature in-plane penetration depth $\lambda_{ab} (0)$ was measured 
to be 
2500 \AA~\cite{Prozorov}. This crystal shows $T^{onset}_{c}$ at 22 K (defined by the onset of diamagnetism)$,$ and 
$T^{mid}_{c}$ at 19 K (defined as the inflection point on 
$\lambda_{ab} (T)$) \cite{Prozorov}. A wide 
superconducting transition in this crystal manifests a rather low quality of 
the crystal.

We fit the data by Eq.~\ref{Se8} with two fitting parameters $\Delta (0)$ and 
$\lambda^{L}_{ab}(0)$$,$ and with a fixed $T_{c}$ = 
20.5 K (the average of 
$T^{onset}_{c}$ and $T^{mid}_{c}$). The solid line is the fitted curve by 
Eq.~\ref{Se8}. It 
is remarkable that the fit is very good. This can be seen more 
clearly in Fig.~\ref{SF7}b where the difference between the data and the 
fitted curve is plotted. There is a negligible systematic error (the 
deviation is less than the magnitude of the data scattering). From 
the fit$,$ we find $\Delta (0)/k_{B}$ = 29.6$\pm$0.1 K$,$ and 
$\lambda^{L}_{ab}(0)$ = 1643 \AA. The deduced $\lambda^{L}_{ab}(0)$ is in 
excellent agreement with the value (1600$\pm$100 \AA) obtained from 
the optical data \cite{Homes}. The magnitude of $2\Delta (0)/k_{B}T_{c}$ = 
2.9 is also the same as the one deduced from a high-quality film where 
the $T^{2}$ term is absent \cite{Alff}. The value of 
$\Delta (0)$ justifies the fit to the data below 6 K$,$ namely$,$ 
0.2$\Delta (0)/k_{B}$. Therefore$,$ the 
$\lambda_{ab} (T)$ data for the crystal are in quantitative agreement 
with an anisotropic $s$-wave pairing symmetry with no nodes.

From the values of $\lambda^{L}_{ab}(0)$ and $\lambda_{ab} (0)$$,$ we calculate 
$\lambda_{ab}^{p}(0)$ = 1884 \AA. Using the relation $\lambda_{ab}^{p}(0)$ = 
$[\Phi_{0}/H_{c}(0)]\sqrt{2N_{d}/\pi}$ and $H_{c}(0)$ = 
2 kOe \cite{Wu}$,$ we estimate $N_{d}$ = 5.2$\times$10$^{10}$/cm$^{2}$$,$ 
corresponding to one defect over 1333 Cu sites. This implies that a 
small density of defects can produce a quite large $\lambda_{ab}^{p}(0)$ which 
contributes a substantial $T^{2}$ term in $\lambda_{ab} (T)$.

\begin{figure}[htb]
    \ForceWidth{7cm}
	\centerline{\BoxedEPSF{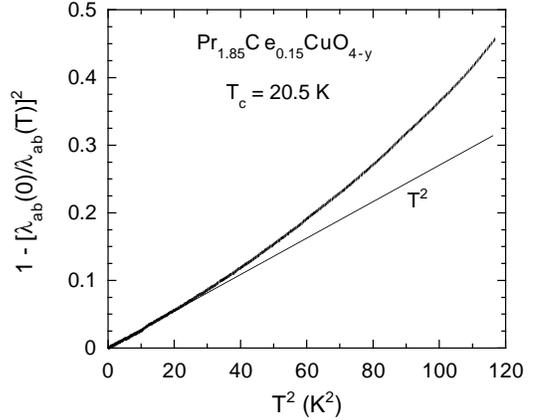}}
	\vspace{0.3cm}
	\caption[~]{The $T^{2}$ dependence of the quantity 
	1-$\lambda_{ab}^{2}(0)/\lambda_{ab}^{2}(T)$ over 0.4$-$10.8 K for the same 
PCCO crystal as the one in Fig.~\ref{SF7}. The crossover from the $T^{2}$ to a 
higher power-law dependence starts at about 5 K. There is no crossover from the $T^{2}$ to the T dependence at $T^{*}$ $\simeq$ 9 K. }
	\protect\label{SF8}
\end{figure}

In order to rule out the possibility that the data can be also consistent 
with a $d$-wave symmetry in the dirty limit$,$ we plot the data 
as $1-\lambda_{ab}^{2}(0)/\lambda_{ab}^{2}(T)$ vs $T^{2}$ in Fig.~\ref{SF8}. It is apparent that 
the quantity $1-\lambda_{ab}^{2}(0)/\lambda_{ab}^{2}(T)$ is proportional to $T^{2}$ 
below about 5 K. For a dirty $d$-wave superconductor$,$ a crossover from 
$T^{2}$ to $T$ dependence should be seen at a temperature $T^{*} 
\simeq \lambda_{ab} (0)\ln 2/[\Delta_{M}(0)d \lambda_{ab} /dT^{2}]$$,$ where 
$\Delta_{M}(0)$ is the maximum gap at zero temperature ~\cite{Hirschfeld}. 
Using $\lambda_{ab} 
(0)$ = 2500 \AA \cite{Prozorov}$,$ $d\lambda_{ab} /dT^{2}$ = 3.7 \AA/K$^{2}$ 
\cite{Prozorov}$,$ and $\Delta_{M}(0)$ = 2.5$T_{c}$ \cite{Stadlober}$,$ one 
finds $T^{*}$ $\simeq$ 9 K.
There is no such a crossover at any temperatures up to 11 K (see 
Fig.~\ref{SF8}). Only a possible crossover from the $T^{2}$ to a 
higher power-law dependence is seen at about 5 K. Therefore$,$ the data cannot 
agree with the $d$-wave pairing symmetry. Furthermore$,$ the absence of 
the linear $T$ term in $\lambda_{ab} (T)$ indicates that the extrinsic 
contribution to $\lambda_{ab} (T)$ due to weak links \cite{HalbJAP} is negligible 
in this crystal.

\begin{figure}[htb]
    \ForceWidth{7cm}
	\centerline{\BoxedEPSF{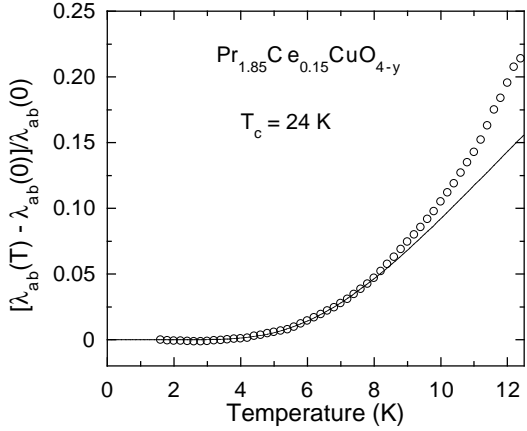}}
	\vspace{0.3cm}
	\caption[~]{Temperature dependence of $[\lambda_{ab} (T) - 
	\lambda_{ab}
(0)]/\lambda_{ab} (0)$ 
	for a high-quality PCCO thin film with the lowest residual 
	resistivity and the highest $T_{c}$. The solid line is the fitted 
	curve by Eq.~\ref{Se9} with $2\Delta (0)/k_{B}T_{c}$ = 2.7. The data are from 
	Ref.~\cite{Alff}. }
	\protect\label{SF9}
\end{figure}

In Fig.~\ref{SF9}$,$ we show 
$[\lambda_{ab} (T) - \lambda_{ab} (0)]/\lambda_{ab} (0)$ as a function of 
temperature for a high-quality 
PCCO thin film (the data are from Ref.~\cite{Alff}). The film has 
the lowest residual resistivity ($<$ 50 $\mu\Omega$ cm) and 
the highest $T_{c}$ (24 K) reported for the PCCO system \cite{Alff}. This 
indicates a high-quality of the film$,$ which was grown using molecular 
beam epitaxy. The optimal quality of the film may be due to the fact 
that a homogeneous oxygen reduction can be 
easily achieved in thin films.  Since the data at low 
temperatures are quite flat$,$ it appears that there is neither $T^{2}$ 
nor $T$ contribution. We thus fit the data below 6.5 K by Eq.~\ref{Se9} with one fitting 
parameter $\Delta (0)$. The best bit gives $\Delta (0)/k_{B}$ = 
31.9$\pm$0.1 K$,$ which justifies the fit to the data below 6.5 K ($\sim$ 
0.2$\Delta (0)/k_{B}$).  This leads to $2\Delta (0)/k_{B}T_{c}$ = 2.7$,$ 
which is nearly the same as that deduced above for the less ideal 
crystal where there 
is a significant $T^{2}$ term in $\lambda_{ab}(T)$ due to the existence 
of defects. All these results consistently suggest that the pairing 
symmetry in electron-doped cuprates is the anisotropic $s$-wave with no line nodes.

Polarized Raman scattering \cite{Stadlober} has also 
shown that the symmetry of the order parameter in 
Nd$_{1.84}$Ce$_{0.16}$CuO$_{4-y}$ (NCCO) is consistent with an anisotropic 
$s$-wave. More precisely$,$ the tunneling spectra \cite{Kashi} are consistent with a 
gap function: $\Delta (\theta ) = \Delta (s +\cos 4\theta)$ with 
$s$ $>$ 1.  If we use $\Delta_{M}$ = 2.5$T_{c}$ 
\cite{Stadlober} and the minimum gap $\Delta_{m}$ = 
1.4$T_{c}$ (from the $\lambda_{ab}(T)$ data)$,$ we find 
$\Delta (\theta ) =  1.15 (3.52 +\cos 4\theta)$ meV for an 
electron-doped cuprate with $T_{c}$ = 24 K.  Therefore$,$ three  bulk-sensitive 
experiments consistently indicate an anisotropic $s$-wave pairing symmetry in 
electron-doped cuprates.

\section{Phase-sensitive experiments along the c-axis direction}

The most reliable phase-sensitive experiment is the atomically clean BSCCO 
Josephson junctions between identical single crystal cleaves stacked 
and twisted an angle $\phi_{0}$ about the c-axis \cite{Li}. The quality of the junction is nearly the 
same as that of the intrinsic Josephson junctions made of the Bi-O 
insulating layers. Theoretically$,$ it has been shown that the critical 
current $I_{c}$ of the twist junction is \cite{Klemm}
\begin{equation}\label{Se10}
I_{c}\propto \sum_{\ell}\eta_{\ell}\Delta_{l}\cos \ell\phi_{0},
\end{equation}
where $\ell$ = 0$,$ 1$,$ 2$,$\ldots.$,$ and $\eta_{\ell}$ $<$$<$ $\eta_{0}$ 
for $\ell$$\geq$1. The above equation indicates that the $s$-wave component 
contributes to the critical current much more effectively. The 
experiment shows \cite{Li} that the $I_{c}$ value is nearly independent of the 
twist angle $\phi_{0}$$,$ and the temperature dependence of $I_{c}$ is consistent with the 
Ambegaokar-Baratoff (AB) model for an $s$-wave superconductor. This indicates 
that the $s$-wave component in this material must 
be significant compared with the other high angular 
momentum components. For slightly overdoped BSCCO$,$ we have found that the 
gap function is $\Delta (\theta ) = 17.75 (\cos 4\theta + 0.46)$ meV for 
$T_{c}$ = 90 K$,$ and $\Delta (\theta ) = 18 (\cos 4\theta + 0.33)$ meV for 
$T_{c}$ = 86 K. Then we have $\Delta_{s}$ = 6-8 meV$,$ which is not 
small compared with the g-wave component $\Delta_{g}$ = 18 meV. Since 
$\eta_{4}$ $<$$<$ 
$\eta_{0}$ \cite{Klemm}$,$ the dominant contribution to the $I_{c}$ 
should be the $s$-wave component$,$ as observed \cite{Li}. From the magnitude of the 
$s$-wave component$,$ we can calculate $I_{c}R_{N}$ = 
$(\pi/2e)\Delta_{s}$ = 9-12 mV. The measured  $I_{c}R_{N}$ value is about 8 
mV \cite{Li}. This is in quantitative agreement with the predicted value considering the 
fact that the strong coupling effect can reduce the $I_{c}R_{N}$ value by more 
than 20$\%$.

Another reliable phase-sensitive experiment is the c-axis 
Pb/YBa$_{2}$Cu$_{3}$O$_{7-y}$ Josephson junction experiment \cite{Sun}. 
This junction can be described as SINS' (where S = YBCO$,$ S'= Pb$,$ I and 
N represent the insulating and normal-metal layers$,$ respectively). Due to 
a very short coherent length $\xi_{c}$ along the c-axis direction$,$ 
the bulk gap will be strongly depressed at the SI interface; the 
depression factor is  $c/\xi_{c}$ (where $c$ is the lattice 
constant along the c-axis) \cite{Muller}. From $\xi_{c} = \xi_{ab}/\gamma$ (where 
$\gamma$ is the mass anisotropy parameter and equal to about 8 for 
optimally-doped YBCO \cite{Willemin2})$,$ we get $\xi_{c}$ = 1.7 \AA~ by taking 
$\xi_{ab}$ = 14 \AA. Therefore$,$ the gap size at SI interface will be 
suppressed by a factor of about 7. Since the bulk $s$-wave component 
$\Delta_{s}$ in slightly overdoped YBCO is 3-5 meV (see above)$,$ this component at 
the SI interface should be reduced to 0.4-0.7 meV. Then the $I_{c}R_{N}$ value 
is calculated to be 0.93-1.27 mV$,$ in quantitative agreement with the 
measured one ($\sim$0.9 mV) \cite{Sun}.

Now we discuss another c-axis Josephson tunneling experiments in 
which a conventional superconductor (Pb) is deposited across a 
single twin boundary of a YBCO crystal \cite{Kou}. Because Pb is an $s$-wave 
superconductor$,$ the Pb counterelectrode couples only to the $s$-wave 
component of the YBCO order parameter. If YBCO were predominantly $d$-wave$,$ any 
small $s$-wave component added to the dominant $d$-wave component would change sign across the twin boundary. 
In this case$,$ magnetic fields parallel to the boundary would produce a 
local minimum in $I_{c}$ at $B$ = 0$,$ in agreement with the 
observation \cite{Kou}. The experimental results thus appear to provide 
evidence for mixed $d$- and $s$-wave pairing symmetry in the bulk with a reversal in the sign of the $s$-wave component 
across the boundary. However$,$ if the bulk OP symmetry in a single domain 
were $d + s$ or $d - s$$,$ one would expect a nearly zero 
$I_{c}$ in heavily twinned crystals. The fact that the observed 
$I_{c}R_{N}$ 
in heavily twinned crystals \cite{Sun} is nearly the same as the one in the 
single-domain 
crystal \cite{Kou} rules out the bulk $d + s$ or $d - s$ wave OP symmetry in YBCO. 
Therefore the only possibility is that a half/fractional flux is trapped in the twin 
boundary. Also$,$ this 
can naturally explain why $I_{c}$ does not go to zero even for a symmetric 
junction with the same junction area in both sides of the twin 
boundary \cite{Kou}.
  
\section{Phase-sensitive experiments along the ab-planes}

The phase-sensitive tricrystal experiments on both hole and electron-doped cuprates \cite{Tsuei1,Tsuei2,Tsuei3} show 
that the OP symmetry is the $d$-wave$,$ in 
contradiction with the above conclusion drawn from many bulk-sensitive 
experiments. In order to resolve the above discrepancy$,$ one should notice that the 
tricrystal experiments are rather 
surface sensitive$,$ so these experiments are probing the OP symmetry 
at the surface/interface$,$ rather than in the bulk. 
Based on the Ginzburg-Landau free energy$,$ Bahcall \cite{Bahcall} has shown 
that the OP symmetry near 
surfaces/interfaces can be different from that in the bulk if the bulk 
OP is strongly suppressed at the surfaces. 
Experimentally$,$ the observed $I_{c}R_{N}$ values in all the 
tricrystal experiments are  about two order of 
magnitude smaller than the intrinsic Ambegaokar-Baratoff limit.  For example$,$ in the 
optimally-doped YBa$_{2}$Cu$_{3}$O$_{7-y}$$,$ the magnitude of the maximum 
gap $\Delta_{M}(0)$ is about 30 meV \cite{Wei,Pon}. Then the 
intrinsic $I_{c}R_{N}$ value should be equal 
to the Ambegaokar-Baratoff limit  $\pi\Delta_{M}(0)/2e$ = 47 mV$,$ which has 
been confirmed by a nearly ideal SIS break junction experiment \cite{Pon}. However$,$ 
the observed  $I_{c}R_{N}$ 
values in the tricrystal experiments 
on YBa$_{2}$Cu$_{3}$O$_{7-y}$ and Tl$_{2}$Ba$_{2}$CuO$_{6+y}$ 
\cite{Tsuei2,Tsuei3} are about 1.8 mV and 0.5 mV$,$ respectively. These values 
are about two order of magnitude smaller than the 
intrinsic bulk values. Similarly$,$ the observed $I_{c}R_{N}$ value in the NCCO 
and PCCO tricrystal experiments is about 0.1 mV$,$ as inferred from the 
measured critical current density $J_{c}$ = 6 A/cm$^{2}$ \cite{Tsuei1} and the 
empirical relation between $I_{c}R_{N}$ and $J_{c}$ \cite{Kleefisch}. This 
$I_{c}R_{N}$ value is also 
about two order of magnitude smaller than the 
intrinsic bulk value$,$ which is estimated to be $\sim$ 8 mV with 
$\Delta_{M}(0)$ = 2.5$T_{c}$ \cite{Stadlober}. Therefore$,$ the OP 
at the interfaces of the 
grain boundary junctions must be strongly depressed in order to 
explain such small $I_{c}R_{N}$ values. This strong depression in the order 
parameter ensures the condition under 
which the OP symmetry near 
surfaces/interfaces can be different from that in the bulk 
\cite{Bahcall}. Hence$,$ it 
is very likely that the 
tricrystal experiments are detecting the OP symmetry at 
the degraded interfaces$,$ which may be different from 
the intrinsic one in the bulk. 

Now the question arises: why the bulk OP can be so 
strongly depressed at the surfaces of the grain boundary junctions? It is known 
that the coherent length 
in cuprates is very short due to a large superconducting gap and small Fermi 
velocity. The short coherent length in cuprates can lead to a large 
depression of the OP near the interfaces even within the 
conventional theory of the proximity effect \cite{Muller,Yu}. Alternatively$,$ 
several groups \cite{Halb1,Bet95,Gurevich} showed that there are 
possibly nonsuperconducting 
regions near the boundary of the junction due to hole depletion 
and/or strain$,$ so that the critical current density can be reduced by 
several order of magnitude compared with the intrinsic bulk value.

There is another way to explain the tricrystal experiments. As 
discussed above$,$ the bounaries of the grain-boundary junctions are 
intrinsically underdoped superconductors or nonsuperconducting 
antiferromagnets due to hole depletion 
and/or strain \cite{Halb1,Bet95,Gurevich,Mann}. For underdoped 
cuprates$,$ the superconductivity mainly arises from the Bose-Einstein 
condensation of preformed pairs \cite{Alexandrov98}. In this case$,$ the 
symmetry of the 
superconducting condensate is different from the pairing symmetry; the 
former is the $d$-wave while the latter might be $s$-wave \cite{Alexandrov98}. 
Since Josephson tunneling probes the symmetry of the superconducting 
condensate$,$ the $d$-wave symmetry of the condensate is consistent with 
the tricrystal experiments.

\section{Conclusion}

In conclusion$,$ the existing bulk and nearly bulk-sensitive experiments 
consistently show that the dominant bulk pairing symmetry in hole-doped 
cuprates is of extended $s$-wave with eight line nodes$,$ and of 
anisotropic $s$-wave in 
electron-doped cuprates. The deduced extended $s$-wave pairing symmetry 
for hole-doped cuprates is also in quantitative agreement with the 
phase-sensitive experiments along the c-axis for both Bi$_{2}$Sr$_{2}$CaCu$_{2}$O$_{8+y}$ 
and YBa$_{2}$Cu$_{3}$O$_{7-y}$. The proposed pairing symmetries do 
not contradict some surface- and phase-sensitive 
experiments which show a predominant $d$-wave pairing symmetry at the 
degraded surfaces.

~\\
~\\
~\\
{\bf Acknowlegement}: The author would like to thank R. Prozorov$,$ L. Alff$,$ S. 
Kamal and W. N. Hardy for sending their 
published data.

\end{document}